\documentclass[aps,prl,twocolumn,superscriptaddress,showpacs,]{revtex4-1}

\usepackage{amsmath}
\usepackage{amssymb}
\usepackage{graphics}
\usepackage{graphicx}
\usepackage{epstopdf}
\usepackage{dcolumn}
\usepackage{bm}
\usepackage{xcolor}

\newcommand{\beq}{\begin{equation}}
\newcommand{\eeq}{\end{equation}}
\newcommand{\bary}{\begin{array}}
\newcommand{\eary}{\end{array}}
\newcommand{\beqary}{\begin{eqnarray}}
\newcommand{\eeqary}{\end{eqnarray}}
\newcommand{\bc}{\begin{center}}
\newcommand{\ec}{\end{center}}
\newcommand{\lgl}{{\langle}}
\newcommand{\rgl}{{\rangle}}
\newcommand{\mtN}{\mathcal{N}}
\newcommand{\mtP}{\mathcal{P}}

\newcommand{\half}{\textstyle \frac{1}{2}}%
\newcommand{\ihalf}{\textstyle \frac{i\beta_2}{2}}%
\begin{document}

\title{Temporal boundary solitons and extreme super-thermal light statistics}

\author{Chunhao Liang}
\affiliation{Shandong Provincial Engineering and Technical Center of Light Manipulations \& Shandong Provincial Key Laboratory of Optics and Photonics Devices, School of Physics and Electronics, Shandong Normal University, Jinan 250014, China}
\affiliation{Department of Electrical and Computer Engineering, Dalhousie University, Halifax, Nova Scotia, B3J 2X4, Canada}
\author{Sergey A. Ponomarenko}
\email[]{serpo@dal.ca}
\affiliation{Department of Electrical and Computer Engineering, Dalhousie University, Halifax, Nova Scotia, B3J 2X4, Canada}
\affiliation{Department of Physics and Atmospheric Science, Dalhousie University, Halifax, Nova Scotia, B3H 4R2, Canada}
\author{Fei Wang}
\affiliation{School of Physical Science and Technology, Soochow University, Suzhou 215006, China}
\author{Yangjian Cai$^{1,}$}
\affiliation{School of Physical Science and Technology, Soochow University, Suzhou 215006, China}

\date{\today}

\begin{abstract}
We discover the formation of a temporal boundary soliton (TBS) in the close proximity of a temporal boundary, moving in a nonlinear optical medium, upon high-intensity pulse collision with the boundary. We show that the TBS excitation causes giant intensity fluctuations in reflection (transmission) from (through) the temporal boundary even for very modest input pulse intensity fluctuations. We advance a statistical theory of the phenomenon and show that the TBS emerges as an extremely rare event  in a nonintegrable nonlinear system, heralded by colossal intensity fluctuations with unprecedented magnitudes of the normalized intensity autocorrelation function  of the reflected/transmitted pulse ensemble. \end{abstract}
\maketitle

Refraction and reflection of waves at a spatial interface separating two media with different refractive indices is a venerable subject. Due to the broken spatial translation symmetry, the spatial wave vector of a monochromatic wave, transmitted through the interface, changes, whereas the wave frequency remains the same. The picture is ubiquitous to waves of any physical nature, such as electromagnetic~\cite{BW} or acoustic~\cite{MI} waves. At the same time, the concept of time refraction, originally introduced in the context of photon acceleration in plasma physics~\cite{Mend01}, has only relatively recently percolated into optics and led to research into the fundamentals of the electromagnetic wave reflection and refraction at a temporal boundary~\cite{Mend02,GPA14,GPA15} and optical pulse transmission through time-varying media~\cite{GPA11,GPA16,GM20}. 

Any temporal boundary (TB) breaks the time-translation symmetry, causing a frequency change upon light transmission through the TB, hence the term photon acceleration. The light behavior at the TB, and in time-varying media in general, can be exploited to explore a multitude of fascinating fundamental phenomena in time domain, such as temporal total internal reflection~\cite{GPA15}, optical nonreciprocity~\cite{Fan09,Alu17}, negative refraction~\cite{Pen08}, photonic topological insulators~\cite{Fan12}, photonic time crystals~\cite{Seg18}, and time reversal~\cite{Fan04}; it can also open up a door to promising applications, including temporal waveguiding~\cite{GPA16,Wu16}, frequency conversion~\cite{Boyd20}, all-optical signal processing~\cite{Vezz18} and reconfigurable photonics~\cite{Alu18,Zhang18}. In the nonlinear optics context, the weak probe pulse collision with a TB, generated by a strong pump pulse, for example, has been instrumental to simulate kinematical effects in gravitational physics~\cite{Leo08,Fac11,Fac12} as well as to study soliton-TB  interactions~\cite{Leo10,Stein11,Stein12,Leo12}. 

To the best of our knowledge, however, only the phenomena arising from bulk optical soliton interactions with the TB have been examined to date. In this Letter, we establish through extensive numerical simulations supported by asymptotic analysis that a {\it temporal boundary soliton} (TBS) can form in the vicinity of a TB moving in a nonlinear optical medium as an optical pulse of sufficiently high intensity, propagating in the medium, interacts with the TB. We further show that if a statistical ensemble of short optical pulses is impinged on such a TB,  the TBS emerges as an extreme random event causing enormous pulse intensity fluctuations, highly localized in time in the TB vicinity, in a given observation plane $z=const$. We demonstrate that such anomalous fluctuations are quantified by unprecedented magnitudes of the normalized intensity autocorrelation function of a reflected/transmitted pulse, exceeding that of thermal light by three orders of magnitude, even for the incident pulse ensembles with fairly small (Gaussian) fluctuations around their averages. Our results open up a new area of research into temporal boundary solitons in time-varying media and establish a link between the two distinct research fields: (temporal) surface soliton theory and the physics of extreme events that has recently witnessed a flurry of research activity aiming to elucidate the fundamental physics behind the rogue wave formation~\cite{RW-rev1,RW-rev2}.

We begin by examining optical pulse propagation in a generic weakly nonlinear medium with weak anomalous dispersion and a time-varying linear refractive index $n(t)$. We assume, for simplicity, that the time variation is encapsulated by a constant refractive index jump $\Delta n$ propagating inside the medium at a constant speed $v_b$, that is, $n(t)=n_0+\Delta n\,\theta(t-z/v_b-t_b)$, where $\theta(x)$ is a Heaviside unit step function taking the value zero at $x<0$ and unity at $x>0$, and $n_0$ is a background linear refractive index of the medium. The electric field $E(t,z)$ of a linearly polarized pulse can be expressed in terms of the slowly-varying envelope $\Psi(t,z)$ and the carrier at frequency $\omega_0$ as $E(t,z)=\Psi (t,z) e^{i(k_0 z-\omega_0 t)}$, where $k_0=n_0 \omega_0/c$ and we assume that  $\Delta n\ll n_0$. In this approximation~\cite{Agra}, the envelope evolution in the medium in the reference frame moving with the TB is governed by the modified nonlinear Schr\"{o}dinger equation (NLSE) in the form
	\beq\label{NLSE}
		\partial_z \tilde{\Psi}+\ihalf\partial_{\tau\tau}^2 \tilde{\Psi}-i k_0\Delta n \theta(\tau-t_b)\tilde{\Psi}-i\gamma|\tilde{\Psi}|^2 \tilde{\Psi}=0.
			\eeq
Here $\tau=t-z/v_b$; $\beta_2<0$ and $\gamma$ are anomalous group velocity dispersion and Kerr nonlinearity coefficients, respectively. We also introduced a gauge transformation viz., $\Psi=\tilde{\Psi}e^{i\nu\tau}$ where $\nu=-\Delta\beta_1/|\beta_2|$, $\Delta\beta_1=\beta_1-v_b^{-1}$ being an inverse group velocity mismatch between the pulse and TB, to remove a convective term associated with the group velocity mismatch.

We performed extensive numerical simulations solving Eq.~(\ref{NLSE}) with a Gaussian input pulse, $\Psi (t,0)=\sqrt{P_0}e^{-t^2/2t_p^2}$, having a fixed width $t_p=0.6$ ps and variable input peak power $P_0$. In all our simulations we chose the group-velocity dispersion and nonlinearity parameters to be generic of a standard silica-glass fiber at the telecommunication wavelength, $\lambda_0 =1.55 \mu$m: $\beta_2=-0.02$ ps$^2$/m and $\gamma=0.002$ W$^{-1}$m$^{-1}$. Further, we selected the TB parameters, $\Delta n=-0.787\times 10^{-7}$ and $\Delta\beta_1=0.1$ ps/m, such that the known condition for the total internal reflection of the pulse at the TB with $t_b=5$ ps in a linear medium, corresponding to $\gamma=0$, is satisfied~\cite{GPA15}. 

\begin{figure}[t]
\centering
   \includegraphics[width=\linewidth]{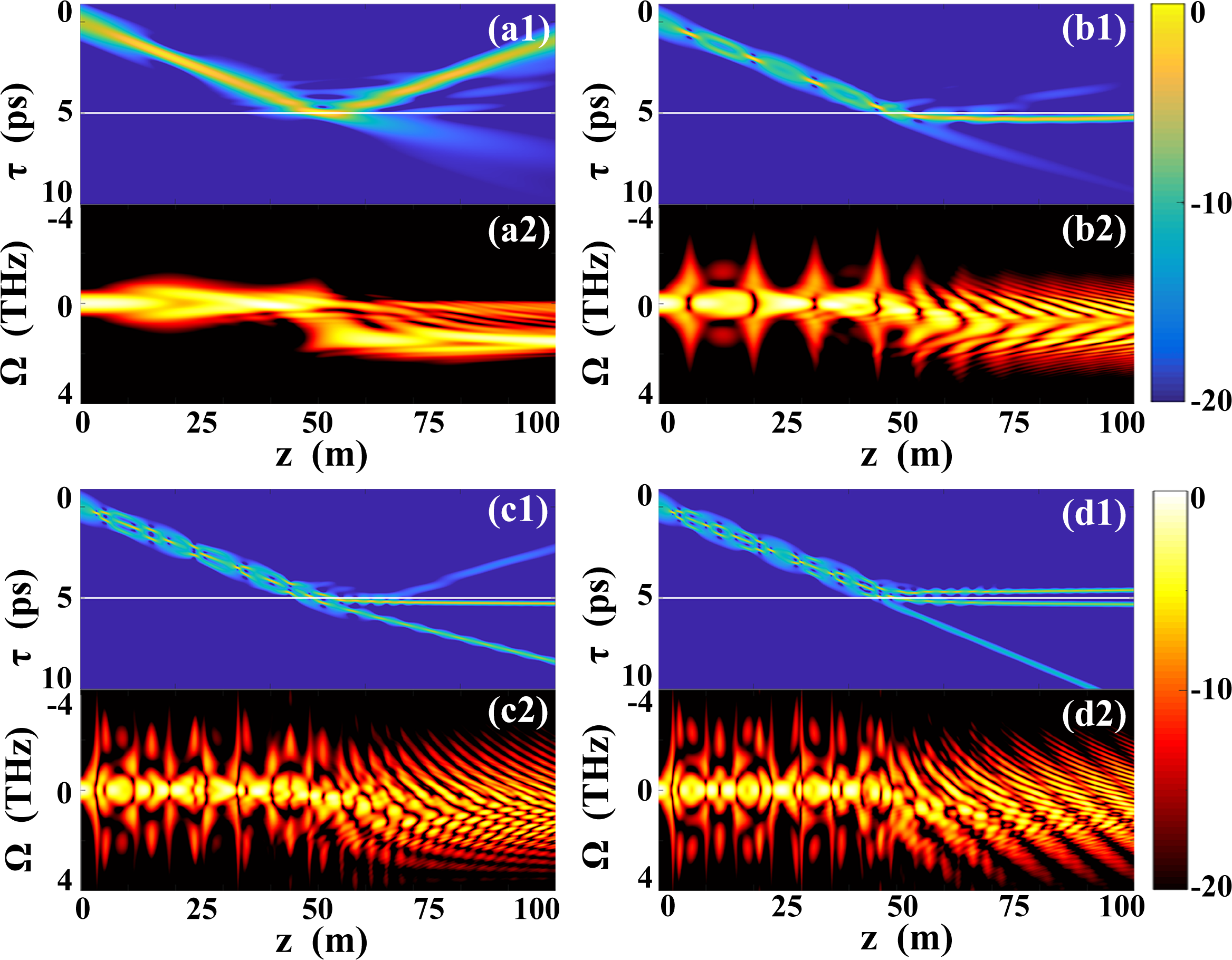}
   \caption{Evolution of the intensity (a1--d1) and the spectrum (a2--b2) of an incident Gaussian pulse of width $t_p=0.6$ ps and input peak power $P_0$: $P_0=80$  W (a1,a2), $P_0=250$ W, (b1,b2) $P_0=593.8$ W, (c1,c2), and $P_0=709.4$ W (d1,d2) as functions of the propagation distance $z$ in the reference frame moving with the temporal boundary as the pulse propagates inside the nonlinear medium with the temporal boundary at $t_b=5$ ps. The intensity and spectral distributions are displayed as functions of the shifted time $\tau=t-z/v_b$ in the TB reference frame and the frequency $\Omega=\omega-\omega_0$ relative to the carrier frequency $\omega_0$ corresponding to the communication wavelength $\lambda=1.55 \mu$m of a standard silica-glass fiber. }
 \label{TBS}
\end{figure}

We exhibit representative results of our simulations in Fig.~\ref{TBS}. As long as the input pulse power $P_0$ is relatively low, $P_0=80$W, a soliton pulse travelling in a cloud of radiated waves---which exist due to non-integrability of Eq.~(\ref{NLSE})---suffers (nearly) total internal reflection at the TB as most of the incident pulse energy is reflected back into $t<t_b$. This scenario is illustrated in Fig.~\ref{TBS}(a1). As $P_0$ is increased to $P_0=250$W, however,  a multi-soliton complex is formed on propagation of the input pulse toward the TB. The multi-soliton complex suffers fission upon encountering the TB. A  significant energy fraction of the multi-soliton is concentrated near the TB and it self-induces a very deep and narrow waveguide via self-phase modulation. It follows that the trapped energy is transported as a sole mode of this self-induced waveguide as is displayed in Fig.~\ref{TBS}(b1). We support our conjecture with a semi-qualitative asymptotic analysis (see Supplemental Material~\cite{Sup} for details). 

Another unambiguous signature of the localization of the discovered TBSs by the TB is revealed by their spectrum, exhibited in Figs~\ref{TBS}(a2--d2) in terms of the relative frequency $\Omega=\omega-\omega_0$. In the spectral domain, the TBS emergence is heralded by the appearance of distinct interference fringes which are clearly discernible on comparing Figs.~\ref{TBS}(a2) and~\ref{TBS}(b2). We attribute these fringes to spectral domain interference between the TBS and TB due to their close overlap in the time domain.

Further numerical simulations point to extreme sensitivity of TBS formation to the magnitude of the incident power $P_0$. Such a sensitivity is illustrated in Videos 1 \& 2~(Supplemental Material~\cite{Sup}) that show the incident multi-soliton complex fission at the TB in the same $P_0$ range as that in Fig.~\ref{TBS}(b1). We can infer from the videos as well as Fig.~\ref{TBS}(b1) that only in a tiny range of $P_0$ are the conditions favorable to the TBS formation in the vicinity of the TB. In the vast majority of  situations, the fission leads to bulk solitons and radiation waves carrying the energy away from the TB. The interaction between the TBS and the other fission products through the cross-phase modulation, neglected within the framework of our asymptotic analysis of~\cite{Sup}, strongly influences the TBS dynamics along the TB and can trigger instability causing the temporal boundary soliton to detach from the TB and propagate away as a bulk soliton. We note that this interaction is, in general, more complex than generic multi-soliton interactions in homogeneous and inhomogeneous integrable systems~\cite{PSA}. 

As we increase the input peak power of the source pulse even further to $P_0=593.79$ W---see Fig.~\ref{TBS}(c1)---the break-up of an incident multi-soliton complex results in a transmitted soliton, a weaker reflected soliton and a TBS, all three strongly interacting via the cross-phase modulation. Finally, as $P_0$ reaches 709.4 W, two TBSs can be simultaneously trapped in the same self-induced waveguide around the TB as is shown in Fig.~\ref{TBS}(d1) for $\Delta n=-0.787\times 10^{-7}$ and confirmed in Video 4(d) (Supplemental Material~\cite{Sup}) for $\Delta n=-0.804\times 10^{-7}$ and the source peak power in the range (715.8 W, 717.8 W) as well. This situation is quite delicate, however,  as a two-boundary-soliton state,  with the two TBSs symbiotically coexisting within the same self-induced waveguide by the TB, can swiftly disintegrate due to tiny fluctuations in $P_0$.  This two-TBS  break-up scenario is captured in Videos 3 \& 4 (Supplemental Material~\cite{Sup}). Figs.~\ref{TBS}(c2) and~\ref{TBS}(d2) also depict a more complex spectral interference pattern between the TBS, TB and the other soliton fission products in this range of $P_0$. 

\begin{figure}[t]
\centering
   \includegraphics[width=\linewidth]{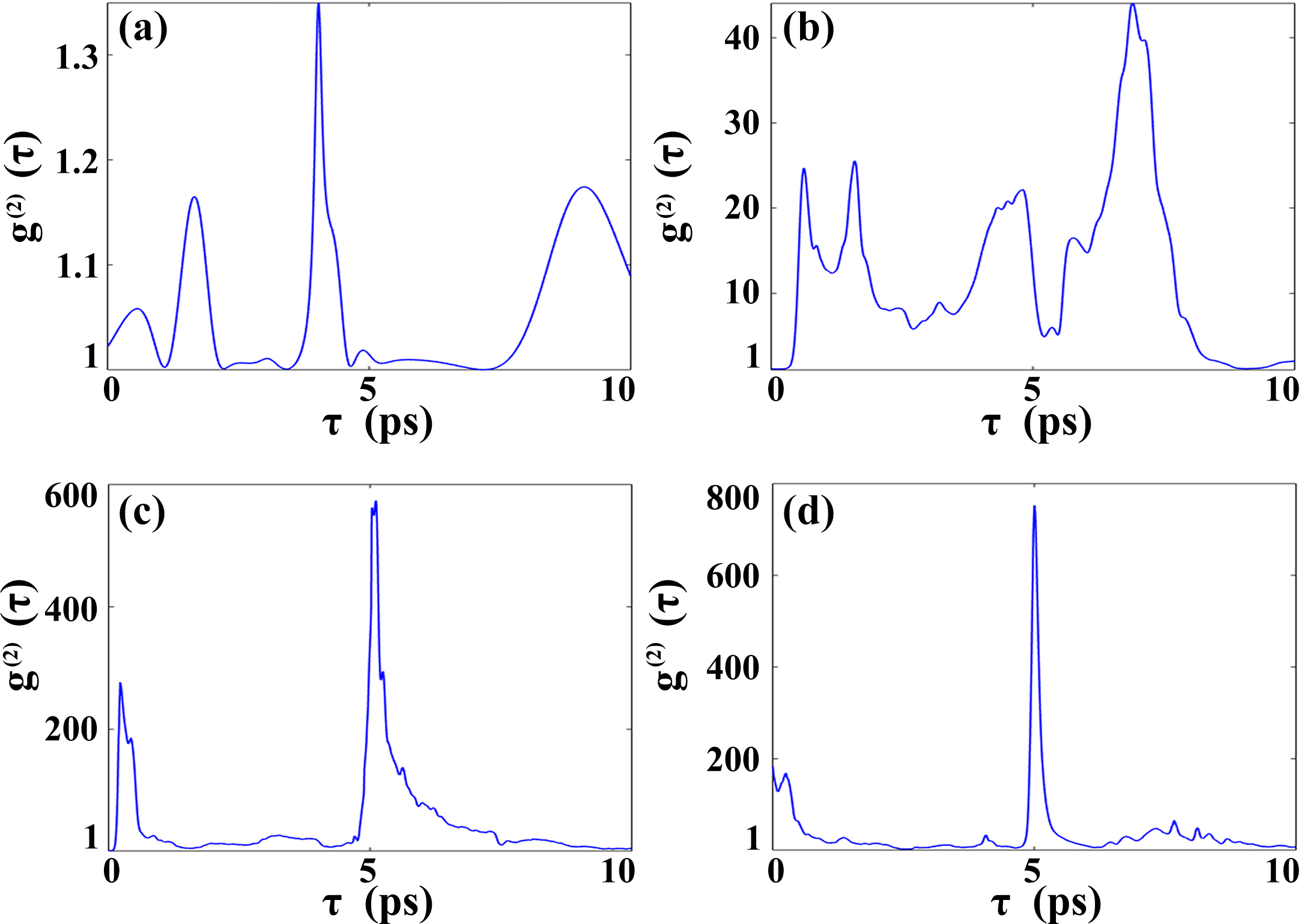}
   \caption{ Normalized intensity autocorrelation function of an ensemble of input Gaussian pulses with the average peak power at the source $\lgl P_0\rgl$: $\lgl P_0\rgl=80$  W (a), $\lgl P_0 \rgl=250$ W (b), $\lgl P_0\rgl=593.8$ W (c), and $\lgl P_0\rgl=709.4$ W (d) as a function of time $\tau$ in the reference frame moving with the TB, evaluated in the receiver plane at $z_{\ast}=100$ m away from the source. }
 \label{g2}
\end{figure}
We have discovered that the TBS generation triggers remarkable statistical anomalies if the source generates random input pulses. To reveal statistical signatures of TBS emergence, we examined an ensemble of 10$^4$ realizations of Gaussian input pulses of fixed width $t_p=0.6$ ps and random input peak power, obeying a Gaussian statistics with the probability density function (PDF) $\mtP(P_0)\propto e^{-(P_0-\lgl P_0\rgl)^2/2\sigma_P^2}$. The PDF is completely determined by the mean peak power $\lgl P_0\rgl$ and variance $\sigma_P$ of the ensemble where the angle brackets denote ensemble averaging. We remark that the Gaussian PDF of the source peak power is quite generic as it corresponds to an approximately Gaussian PDF  of the soliton parameter, provided the fluctuations of $P_0$ are not too large (Supplemental Material~\cite{Sup}). Next,  we constructed statistical ensembles with a given relative noise level in terms of the ratio of $\sigma_P$ to $\lgl P_0\rgl$. The source noise level ranges from 3\%, to 10\%, corresponding to weakly to moderately fluctuating pulses, respectively. We performed Monte-Carlo simulations and evaluated the normalized intensity autocorrelation function, defined as~\cite{MW}
	\beq\label{g2-def}
		g^{(2)}(\tau)=\lgl |\Psi (\tau, z_{\ast})|^4 \rgl/\lgl |\Psi (\tau,z_{\ast})|^2\rgl^2,
			\eeq
in the observation plane at $z_{\ast}=100$ m from the source.  

We exhibit the results in Fig.~\ref{g2} for  weakly fluctuating (3\% noise) ensembles with $\lgl P_0\rgl$  taking on the values 80 W, 250 W, 593.8 W \& 709.4 W, corresponding, for simplicity, to the incident soliton fission scenarios displayed in Fig.~\ref{TBS}. We can infer from the figures that at low $\lgl P_0\rgl$, the global maximum of $g^{(2)}$ is on the order of unity, but as the TBS forms---see Fig.~\ref{TBS}(b1)---the global maximum jumps up by an order of magnitude, reaching the magnitudes in excess of 40 for the ensemble with $\lgl P_0\rgl$=250 W. Thus, a dramatic increase in the magnitude of the maximum of the normalized intensity autocorrelation function signifies the TBS generation. Further increase in the average peak power leads to colossal intensity fluctuations encapsulated by $g^{(2)}$ attaining the values of nearly 600 and 800 in Figs.~\ref{g2}(c) and ~\ref{g2}(d). In addition, as the average peak power of the source exceeds certain threshold, around 600 W in our parameter regime, the shape of $g^{(2)}(\tau)$ changes qualitatively. In particular, a relatively broad $g^{(2)}$ distribution with the main peak past the TB, $\tau_{max}>t_b=5$ ps in Fig.~\ref{g2}(b) is superseded by an extremely tall and narrow spike centered around the TB, $\tau_{\max}\simeq t_b$ and sitting atop a broad background distribution which is evident in Figs.~\ref{g2} (c) and (d).
\begin{figure}[h!]
\centering
   \includegraphics[width=3.0in]{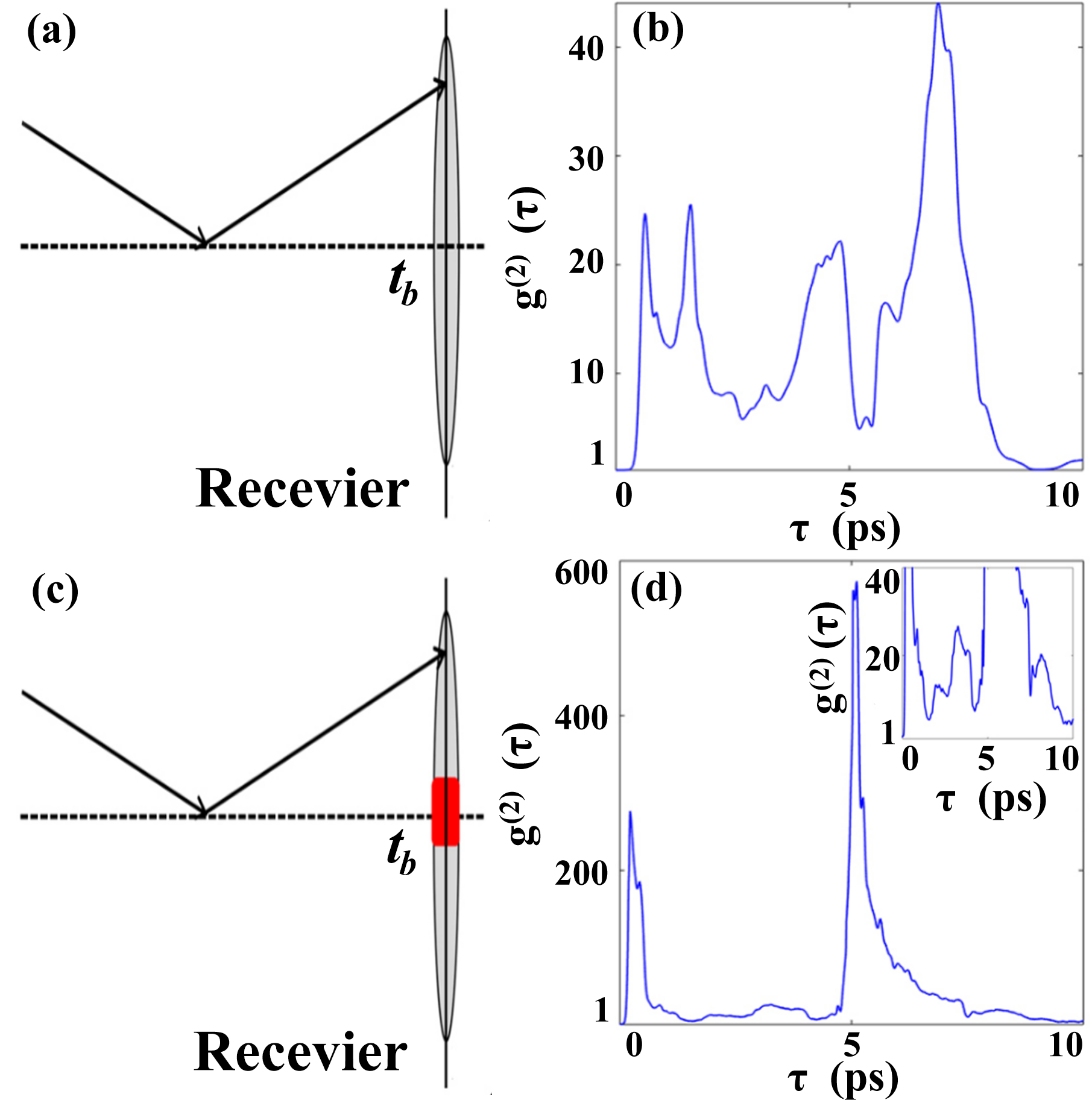}
   \caption{Illustrating the shape of the normalized autocorrelation function. (a) The grey area corresponds to the time interval at the receiver where TBS and/or other soliton fission product are likely to arrive. (b) The broad normalized autocorrelation function $g^{(2)}(\tau)$, corresponding to a pulse ensemble with $\lgl P_0\rgl= 250$ W and 3\% noise level, evaluated at $z_{\ast}=100$, exhibits a main peak and several secondary peaks of comparable magnitude. (c) The red area corresponds to the time interval at the receiver where exceptionally rare TBSs, which propagate all the way along the TB toward the receiver plane, arrive. (d) The normalized intensity autocorrelation function, corresponding to a pulse ensemble with $\lgl P_0\rgl= 593.8$ W and 3\% noise level, evaluated at $z_{\ast}=100$, features a tall and narrow spike residing on a broad pedestal comprised of a multitude of much lower peaks of comparable heights. Inset: A broad pedestal of $g^({2)}$ for the same ensemble as in panel (d) sports multiple peaks of the height of the same order of magnitude as that of the peaks in panel (b). }    
\label{g2qual}
\end{figure}		

We now elucidate the discovered statistical anomalies. First, the colossal intensity fluctuations associated with the TBS inception can be explained by revisiting the behavior of a single ensemble realization. As the realization (pulse) propagates inside the medium, a multi-soliton complex forms, accompanied by some radiation on account of non-integrability of our modified NLSE. After a collision with the TB and subsequent soliton fission, in most cases the repulsion among the fission products inhibits the TBS formation. In these situations, the pulse arriving at the observation plane/receiver has a relatively low peak power. In rare cases, however, the TBS emerges and a very high peak power pulse arrives at the observation plane. Thus, denoting by $N_{\mathrm{eff}}$ $(N_{\mathrm{eff}}\gg 1)$ an effective number of realizations reliably reproducing the source ensemble statistics for a given noise level, we surmise that $\lgl |\Psi|^2\rgl$ and $\lgl |\Psi|^4 \rgl$ at the position in the observation plane where the TBS arrives are dominated by a single ensemble realization $\Psi_{TBS}$ corresponding to the case when the TBS is formed. It then follows at once that $\lgl |\Psi|^2\rgl=N_{\mathrm{eff}}^{-1}\sum_j |\Psi_j|^2\simeq |\Psi_{TBS}|^2/N_{\mathrm{eff}}$ and $\lgl |\Psi|^4\rgl=N_{\mathrm{eff}}^{-1}\sum_j |\Psi_j|^4 \simeq |\Psi_{TBS}|^4/N_{\mathrm{eff}}$. Therefore, Eq.~(\ref{g2-def}) implies that 
	\beq\label{g2-TBS}
		g_{\mathrm{max}}^{(2)}\gtrsim N_{\mathrm{eff}}\gg 1, 
			\eeq
that is, giant magnitudes of the intensity autocorrelation function can be expected even for rather modest source noise levels. 
		
To explain the pronounced growth of the global maximum of $g^{(2)}$ with $\lgl P_0\rgl$ for a given noise level, we recall that given the latter, our ensembles are constructed such that the variance of the source peak power PDF augments with $\lgl P_0\rgl$, making the source effectively noisier in terms of the peak power, and hence calling for a greater number of realizations $N_{\mathrm{eff}}$ to reproduce the source statistics. It then follows from Eq.~(\ref{g2-TBS}) that the global maximum of $g^{(2)}$ must grow with $\lgl P_0\rgl$ as well.
	
Next, we refer to Fig.~\ref{g2qual} from which the structure of $g^{(2)}(\tau)$ can be understood. The broad pedestal on which a huge spike of $g^{(2)}$ is situated for sufficiently large $\lgl P_0\rgl$ corresponds to the time interval of the receiver plane colored in grey in Figs.~\ref{g2qual}(a) and~\ref{g2qual}(c). This is the receiver time window where the TBS is likely to arrive after it has separated from the TB due to the cross-phase modulation induced instability. At the same time, the much shorter time interval, depicted in red in Fig.~\ref{g2qual}(c), contains the window in the neighbourhood of $t_b$ where the TBS can arrive in a very unlikely scenario that it is capable of travelling all the way along the TB toward the receiver, overcoming the overwhelming odds of having being peeled off of the TB through its interaction with the other (high-power soliton) fission products. This is an extremely rare event that can happen only for very high $\lgl P_0\rgl$, which explains the presence of the spike in Fig.~\ref{g2qual}(d) representing the ensemble with $\lgl P_0\rgl=593.8$ W and its absence in Fig.~\ref{g2qual}(b) visualizing the ensemble with $\lgl P_0\rgl=250$ W. Notice that the broad pedestal in Fig.~\ref{g2qual}(d) corresponds to the values of $g^{(2)}$ of the same order as those in Fig.~\ref{g2qual}(b) as is manifested in the inset to Fig.~\ref{g2qual}(d). 
	
\begin{figure}[h!]
\centering
   \includegraphics[width=\linewidth]{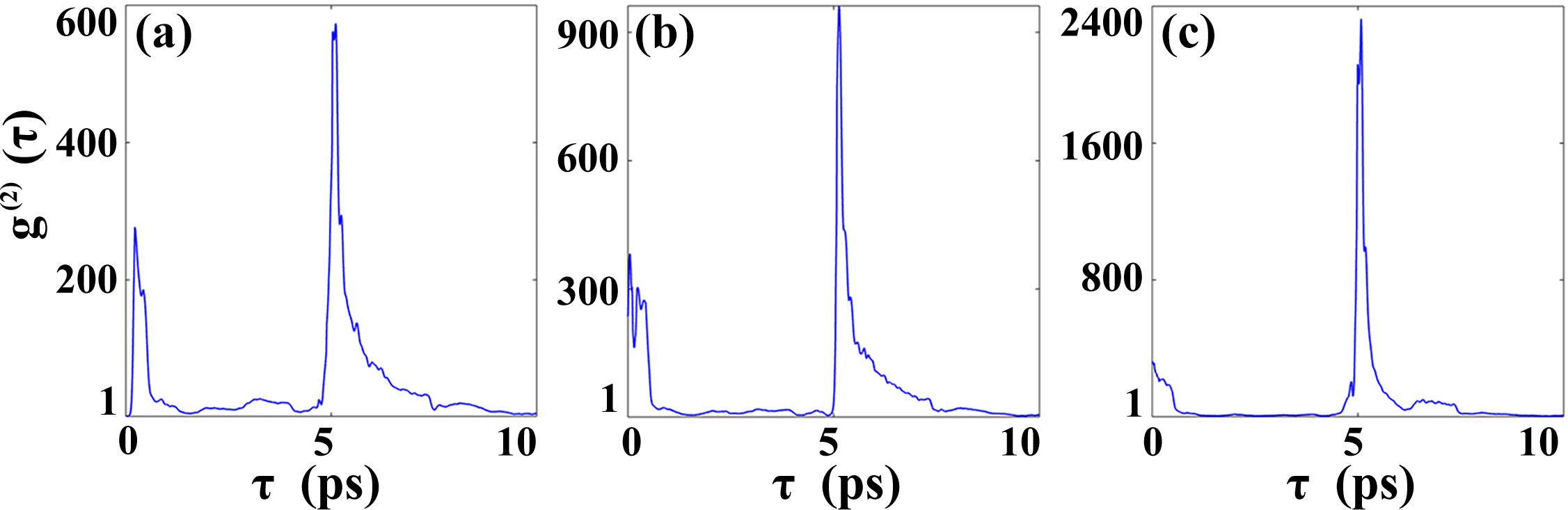}
   \caption{Normalized intensity autocorrelation function of an ensemble of input Gaussian pulses with the average peak power at the source $\lgl P_0\rgl=593.8$  W and 3\% (a), 5\% (b) and 10\% (c) noise level as a function of time $\tau$ in the reference frame moving with the TB, evaluated in the receiver plane at $z_{\ast}=100$ m away from the source.}
   \label{g2-noise}
\end{figure}	
Finally,  we display in Fig.~\ref{g2-noise} the dependence of the intensity autocorrelation function on the noise level of the source for a given (high) average peak power of the input pulse ensemble, $\lgl P_0\rgl=593.8$ W. It can be readily inferred from the figure that $g^{(2)}$ grows sharply as the source noise level is increased, attaining unprecedented, to our knowledge, magnitudes of 962 and 2327 for 5\% and 10\% noise levels, respectively. The revealed unprecedented values of $g^{(2)}$ exceed that of thermal light by more than three orders of magnitude. The precipitous growth of $g^{(2)}$ with the noise level can be qualitatively explained by recalling that one requires a larger number of ensemble realizations to faithfully reproduce a noisier ensemble, which, in turn, implies the increase of the maximum of $g^{(2)}$ by virtue of Eq.~(\ref{g2-TBS}). 

In conclusion, we stress that unlike previous studies concerned with generating bulk solitons upon ultrashort---in a femtosecond range---pulse fission at an optical even horizon~\cite{Leo10,Stein11, Stein12,Leo12} or with guiding weak pulses inside femtosecond soliton-induced waveguides~\cite{OPEX}, we have revealed temporal boundary soliton generation, such that the TBS guides itself along the TB inside a self-induced waveguide in the optical medium. In addition, the observation of the predicted super-thermal light statistics requires only picosecond pulses with modest input peak powers triggering no supercontinuum generation. We expect our results to shed new light on extreme event excitation in media with temporal event horizons, thereby contributing to a thriving multidisciplinary field of rogue wave physics. 

The authors acknowledge financial support from the Natural Sciences and Engineering Research Council of Canada, (RGPIN-2018-05497); the National Natural Science Foundation of China (NSFC) (12004220, 11525418, 91750201, 11974218), China Postdoctoral Science Foundation (2019M662424), the National Key Research and Development Project of China (2019YFA0705000); the Innovation Group of Jinan (2018GXRC010), and the Local Science and Technology Development Project of the Central Government (No. YDZX20203700001766).

	\bc
   {\bf \large Supplemental Material}
   	\ec

{\bf S1. Asymptotic theory of temporal boundary solitons (TBS) in the high-power short-width limit}---Guided by our numerical simulations, we focus on the case of very high peak power input pulses. As a source Gaussian pulse of sufficiently high input power propagates in the medium, a multi-soliton complex forms that suffers fission (break-up) upon encountering the temporal boundary (TB). A  significant energy fraction of the multi-soliton complex  is concentrated near the TB and it self-induces a very deep and narrow waveguide via the self-phase modulation. The TBS then propagates as a {\bf sole} mode of this self-induced waveguide.  We assume the trapped portion of the pulse to have a very high peak power $P_s$, so that the nonlinearity dominates the linear refractive index jump: $\gamma P_s\gg k_0\Delta n$. We stress that although in the following $\Delta n$ never explicitly enters, the magnitude of the linear refractive index jump plays an important role in the multi-soliton complex fission scenario. In particular, it will strongly affect the amount of energy $W_s$ trapped by the TBS, which is a key parameter describing the latter.  In the nonlinearity dominated regime, the modified NLSE can be written as
	\beq\label{SPM-NLSE}
		\partial_{z} \Psi +\Delta\beta_1 \partial_{\tau} \Psi +\ihalf \partial_{\tau\tau}^2 \Psi-i\gamma |\Psi|^2 \Psi=0,\tag{S1}
			\eeq
where we neglected the linear refractive index jump term. We seek a soliton solution to~(\ref{SPM-NLSE}) localized at the TB in the form
	\beq\label{Anz}
		\Psi(\tau,\zeta)=[\overline{\Psi}_< (\tau)\theta(t_b-\tau)+\overline{\Psi}_> (\tau)\theta(\tau-t_b)]e^{i\beta_s z},\tag{S2}
			\eeq
where $\theta(x)$ is a unit step-function, $\beta_s$ is a soliton propagation constant and we can introduce a soliton profile on each side of the TB as
	\beq
		\overline{\Psi}(\tau)=\left\{\bary{cc}
						\overline{\Psi}_< (\tau),  & \tau<t_b; \\
						\overline{\Psi}_> (\tau),   & \tau>t_b.
							\eary
								\right.\tag{S3}
									\eeq
On substituting the Ansatz~(\ref{Anz}) into Eq.~(\ref{SPM-NLSE}), we arrive at
	\beq\label{aux1}
		\overline{\Psi}^{\prime\prime}+\frac{2i\Delta\beta_1}{|\beta_2|}\overline{\Psi}^{\prime}-\frac{2\beta_s}{|\beta_2|}\overline{\Psi} +\frac{2\gamma}{|\beta_2|}|\overline{\Psi}|^2 \overline{\Psi}=0,\tag{S4}
			\eeq
where a prime denotes a derivative with respect to $\tau$. It will prove convenient to get rid of the first derivative term in Eq.~(\ref{aux1}). To this end, we introduce a gauge transformation
	\beq
		\overline{\Psi}=\tilde{\Psi}e^{i\nu\tau},\tag{S5}
			\eeq
into Eq.~(\ref{aux1}) and, provided  $\nu=-\Delta\beta_1/|\beta_2|$, we obtain,
	\beq\label{aux2}
		\tilde{\Psi}^{\prime\prime}-\kappa_s^2\tilde{\Psi} +v(\tau)\tilde{\Psi}=0.\tag{S6}
			\eeq
Here we introduced 
	\beq\label{kap}
		\kappa_s^2=\frac{2\beta_s}{|\beta_2|}-\frac{\Delta\beta_1^2}{|\beta_2|^2},\tag{S7}
			\eeq
and a self-induced ``potential",
	\beq
		v(\tau)=\frac{2\gamma}{|\beta_2|}|\tilde{\Psi}(\tau)|^2.\tag{S8}
			\eeq

Although the actual self-induced  ``potential" is unknown, in the limit of a ``tall" and ``narrow" one, we can draw some important quantitative conclusions without the knowledge of its precise shape. To see this, let us replace the actual potential with a rectangular one. The prior numerical simulations indicate that the TBS lies below the TB, so we replace its actual shape with a rectangular shape, so that
	\beq
		v(\tau)=\left\{\bary{ccc}
						0,  & \tau<t_b; \\
						v_0=2\gamma P_s/|\beta_2|,\,\,\,\,   & t_b < \tau < t_b +t_s; \\
						0, & \tau>t_b+t_s.
							\eary
								\right.\tag{S9}
									\eeq
Here $P_s$ and $t_s$ are the power and width, respectively, of our toy-model rectangular shape soliton. The problem of finding the propagation constant of such a toy model soliton is simplified by shifting the time origin so that the self-induced potential is symmetric with respect to the origin. Accordingly, we introduce a new time variable $\tau^{\prime}$:
	\beq
		\tau^{\prime}=\tau-t_b-t_s/2.\tag{S10}
			\eeq
We can then cast Eq.~(\ref{aux2}) into the form
	\beq\label{aux3}
		\tilde{\Psi}^{\prime\prime}-\kappa_s^2\tilde{\Psi} +v(\tau^{\prime})\tilde{\Psi}=0,\tag{S11}
			\eeq			
where
	\beq\label{v0}
		v(\tau^{\prime})=\left\{\bary{cc}
						v_0=2\gamma P_s/|\beta_2|, \,\,\,\,  & |\tau^{\prime}|<t_s/2;\\
						0, &  |\tau^{\prime}|>t_s/2.
							\eary
								\right.\tag{S12}
									\eeq		
It then follows that
	\beq\label{aux4}
		\tilde{\Psi}_{out}^{\prime\prime} -\kappa_s^2 \tilde{\Psi}_{out}=0, \hspace{1cm} |\tau^{\prime}|>t_s/2,\tag{S13}
			\eeq
describes the soliton profile outside the self-induced waveguide and
	\beq\label{aux5}
		\tilde{\Psi}_{in}^{\prime\prime} +\alpha^2 \tilde{\Psi}_{in}=0, \hspace{1cm} |\tau^{\prime}|<t_s/2,\tag{S14}
			\eeq
does inside the waveguide. Here we introduced
	\beq\label{alf}
		\alpha=\sqrt{v_0-\kappa_s^2}.\tag{S15}
			\eeq
			
Prior to finding the soliton propagation constant as a function of its energy, we observe that as follows from Eq.~(\ref{aux4}) the soliton exists as a bound state provided,
	\beq\label{aux6}
		\kappa_s^2\geq 0.\tag{S16}
			\eeq
It then follows at once from Eqs.~(\ref{kap}) and~(\ref{aux6}) that the soliton propagation constant must satisfy the constraint
\beq\label{TIR}
	 \beta_s\geq \half\Delta\beta_1^2/|\beta_2|.\tag{S17}
		\eeq	
The latter is equivalent to the condition for total internal reflection of a pulse at the TB in a linear medium~\cite{GPA15} with the propagation constant $k_0\Delta n$ related to a linear refractive index jump $\Delta n$. This makes perfect sense as soliton trapping occurs due to self-induced total internal reflection.

The symmetry of the potential $v(-\tau^{\prime})=v(\tau^{\prime})$ suggests that the modes have definite parity: there are even and odd parity modes that can be treated separately. 	\\
{\it Even modes---}We look for solutions to Eqs.~(\ref{aux4}) and~(\ref{aux5}) in the form
	\beq
		\tilde{\Psi}_{out}=Ae^{-\kappa_{even} |\tau^{\prime}|},  \hspace{1cm} |\tau^{\prime}|>t_s/2,\tag{S18}
			\eeq
and
	\beq
		\tilde{\Psi}_{in}=B \cos(\alpha \tau^{\prime}),  \hspace{1cm} |\tau^{\prime}|<t_s/2.\tag{S19}
			\eeq
The field and its derivative must be continuous across $\tau^{\prime}=\pm t_s/2$, implying the continuity of logarithmic derivatives:
	\beq
		\left. \frac{\tilde{\Psi}_{out}^{\prime}}{\tilde{\Psi}_{out}}\right|_{\tau^{\prime}=t_s/2}=\left. \frac{\tilde{\Psi}_{in}^{\prime}}{\tilde{\Psi}_{in}}\right |_{\tau^{\prime}=t_s/2},\tag{S20}
			\eeq
yielding
	\beq\label{even}
		\kappa_{even}=\alpha \tan(\alpha t_s/2).\tag{S21}
			\eeq
{\it Odd modes---}We now look for solutions to Eqs.~(\ref{aux4}) and~(\ref{aux5}) in the form
	\beq
		\tilde{\Psi}_{out}=Ce^{-\kappa_{odd} |\tau^{\prime}|},  \hspace{1cm} |\tau^{\prime}|>t_s/2,\tag{S22}
			\eeq
and
	\beq
		\tilde{\Psi}_{in}=D \sin(\alpha\tau^{\prime}),  \hspace{1cm} |\tau^{\prime}|<t_s/2.\tag{S23}
			\eeq
The continuity of logarithmic derivatives at  $\tau^{\prime}=\pm t_s/2$, yields the equation for the soliton propagation constant
	\beq\label{odd}
		\kappa_{odd}=-\alpha \cot(\alpha t_s/2).\tag{S24}
			\eeq

Let us consider the ultrashort, high-power soliton limit: $v_0\rightarrow\infty$, $t_s\rightarrow 0$,  $v_0 t_s=g_s=const$, corresponding to a finite energy $W_s=P_s t_s$ TBS. We can infer that
$\alpha=\sqrt{v_0^2-\kappa_s^2}\rightarrow \sqrt{v_0}$ as $v_0\rightarrow\infty$. It then follows that $\alpha t_s/2\rightarrow \sqrt{v_0 t_s^2}/2\rightarrow 0$ as $t_s\rightarrow 0$. Further, we can approximate
$\tan(\alpha t_s/2)\simeq \alpha t_s/2$ and $\cot (\alpha t_s/2)\simeq 2/(\alpha t_s)$. The inspection of Eqs.~(\ref{even}) and~(\ref{odd}) reveals that there can be no odd modes and only one even mode, so that
\beq\label{aux6a}
	\kappa_s=\alpha^2 t_s/2 =v_0 t_s/2=g_s/2.\tag{S25}
		\eeq
It follows at once from Eqs.~(\ref{kap}),~(\ref{v0}) and~(\ref{aux6a}) that
\beq\label{betas}
	\beta_s=\frac{\Delta\beta_1^2 +\gamma^2 W_s^2}{2|\beta_2|}.\tag{S26}
		\eeq
		
Thus, the high-power bound state is indeed a TBS that propagates along the TB and exponentially decays away from it. The TBS propagation constant is determined solely by the group velocity mismatch and the overall energy $W_s$ trapped by the TBS, but not by its peak power or temporal width individually. In particular, whenever the TBS energy is low enough, $W_s \ll \Delta\beta_1/\gamma$, the group velocity mismatch alone determines the TBS propagation constant. Moreover, the precise shape of a TBS induced temporal waveguide is largely irrelevant. This observation allows a simple and far reaching generalization of our toy model. Indeed, the key TBS parameters depend only on
\beq
	g_s=\int d\tau^{\prime} v(\tau^{\prime})=\frac{2\gamma}{|\beta_2|} \int d\tau^{\prime}\,|\tilde{\Psi}(\tau^{\prime})|^2,\tag{S27}
		\eeq
which can be expressed in terms of the TBS energy as
\beq\label{gs}
	g_s=2\gamma W_s/|\beta_2|.\tag{S28}
		\eeq
Hence, the high-power ultrashort TBS can be viewed as a sole mode of a self-induced $\delta$-like waveguide, centered at the TB:
\beq\label{TBS-gen}
	\tilde{\Psi}^{\prime\prime}-\kappa_s^2 \tilde{\Psi}+g_s \delta(\tau^{\prime})\tilde{\Psi}=0.\tag{S29}
		\eeq
Indeed, it follows at once from Eq.~(\ref{TBS-gen}) that the TBS profile in this limit can be expressed as
\beq\label{aux7}
	\tilde{\Psi}(\tau^{\prime})=Ae^{-\kappa_s |\tau^{\prime}|}.\tag{S30}
		\eeq
Using the boundary condition for the logarithmic derivative across the self-induced waveguide,
\beq
	\lim_{\epsilon\rightarrow 0} \left. \frac{\tilde{\Psi}^{\prime}}{\tilde{\Psi}}\right|_{\tau^{\prime}=-\epsilon}^{\tau^{\prime}=+\epsilon}=-g_s,\tag{S31}
		\eeq
we readily obtain
\beq\label{aux8}
	\kappa_s=g_s/2,\tag{S32}
		\eeq
Eq.~(\ref{aux8}) implies, by comparing with Eq.~(\ref{aux6a}), the same soliton propagation constant as in Eq.~(\ref{betas}). Eq.~(\ref{aux7}) should be interpreted as follows. In the high-power $P_s \rightarrow\infty$, 
short duration, $t_s\rightarrow 0$, $P_s t_s=const$ limit, the TBS soliton tails $\tau^{\prime}\gg t_s$ are exponential, such that 
\beq
	\Psi (\tau^{\prime}, z)\underset{\tau^{\prime}\rightarrow\pm \infty}{\sim} e^{-g_s |\tau^{\prime}|/2}e^{i\nu\tau^{\prime}}e^{i\beta_s z}.\tag{S33}
		\eeq

At this point, we would like to mention that our asymptotic theory tacitly neglects the interaction between the TBS and the other fission products via cross-phase modulation. This interaction can trigger TBS instability causing the TBS to separate from the TB and propagate at an angle to it; alternatively, the repulsion between a dominant high-power fission soliton and the remaining fission products can preclude the TBS formation at all for certain peak powers of input pulses. Either way, we stress that these scenarios, observed in our numerical simulations, cannot be quantitatively treated within the just presented asymptotic TBS theory framework.   

{\bf S2. Statistical ensemble of the source pulse}---As we are interested in high-power input pulses generating multi-soliton inputs, we fix the pulse width $t_p$ and model the input pulse ensemble as having a high average power $\lgl P_0\rgl$ and weak Gaussian power fluctuations around it. The peak power PDF reads
	\beq\label{Gs}
		\mtP(P)=\frac{1}{\sqrt{2\pi} \sigma_P}\exp\left[-\frac{(P-P_0)^2}{2\sigma_P^2}\right], \tag{S34}
			\eeq
where $\sigma_P\ll P_0$.
			
We note that the pulse evolution is completely determined by two governing parameters: $k_0\Delta n$ and $\mtN= (\gamma P_0 t_p^2/|\beta_2|)^{1/2}$; the latter is known in optics as a soliton parameter~\cite{Agra}. 
We can easily derive the PDF of the soliton parameter $\mtN$ as follows
	\beq\label{N}
		\mtN^2=\frac{\gamma P t_p^2}{|\beta_2|} \,\,\Longrightarrow\,\, dP=2\mtN d\mtN \frac{|\beta_2|}{\gamma t_p^2}.\tag{S35}
			\eeq		
Next, we notice that the probability transformation implies that
	\beq\label{Ptft}
		\mtP(P)dP=\mtP(\mtN) d\mtN.\tag{S36}
		\eeq
It follows from Eqs.~(\ref{Gs}), ~(\ref{N}) and~(\ref{Ptft}) upon elementary algebra that
	\beq\label{PN}
		\mtP(\mtN)=\frac{2\mtN}{\sqrt{2\pi} \sigma_{\mtN}}\exp\left[-\frac{(\mtN^2-\mtN_0^2)^2}{2\sigma_{\mtN}^2}\right] \tag{S37}
			\eeq
is a properly normalized PDF of the soliton parameter with the peak soliton parameter $\mtN_0$ and the soliton parameter dispersion $\sigma_{\mtN}$ defined as
	\beq\label{aux9}
		\mtN_0^2=\frac{\gamma P_0 t_p^2}{|\beta_2|}, \hspace{0.5cm} \sigma_{\mtN}=\frac{\gamma t_p^2}{|\beta_2|}\sigma_P.\tag{S38}
			\eeq
Notice that $\mtP(\mtN)$ is even more sharply peaked than $\mtP(P)$. The former can however be approximated by a Gaussian as can be easily observed from the following:
	\beq\label{aux10}
		\mtN^2-\mtN_0^2=(\mtN-\mtN_0)(\mtN+\mtN_0)\simeq 2\mtN_0 (\mtN-\mtN_0).\tag{S39}
			\eeq
On substituting from Eq.~(\ref{aux10}) into~(\ref{PN}) and replacing the numerator of the prefactor with $2\mtN_0$ at the same level of accuracy, we arrive at
	\beq\label{PN-Gs}
		\mtP(\mtN)=\frac{1}{\sqrt{2\pi} \sigma_{eff}}\exp\left[-\frac{(\mtN-\mtN_0)^2}{2\sigma_{eff}^2}\right],\tag{S40}
			\eeq
where
	\beq\label{aux11}
		\sigma_{eff} =\frac{\sigma_{\mtN}}{2\mtN_0}.\tag{S41}
			\eeq
It then follows at once from Eqs.~(\ref{aux9}) and~(\ref{aux11}) that
	\beq\label{sigeff}
		\sigma_{eff}=\frac{\mtN_0}{2P_0}\sigma_P.\tag{S42}
			\eeq
Thus, $\mtP(\mtN)$ is a quite sharply peaked Gaussian. This completes our source ensemble formulation. 

{\bf S3. Supplementary movies}---We present below a series of videos of the incident soliton-like pulse fission at the TB in the reference frame moving with the TB. In all movies, the incident pulse has the width $t_p=0.6$ ps, and the temporal boundary is located at $t_b=5$ ps. The medium parameters are: $\beta_2=-0.02$ ps$^2$/m, $\gamma=0.002$ W$^{-1}$m$^{-1}$,  and $\Delta\beta_1=0.1$ ps/m. The refractive index jump is equal to  $\Delta n=-0.787\times 10^{-7}$ and $\Delta n=-0.804\times 10^{-7}$ in Movies 1 \& 2 and 3 \& 4, respectively. 

\hspace*{\fill} \\
{\bf Supplementary Movie 1:} This movie shows the evolution of a pulse propagating in the nonlinear optical medium as the input peak power $P_0$ is varied from 210W to 290W. Notice that only in very rare cases, characterized by specific input power levels, can a TBS form. The fission scenario of an input soliton pulse, upon its collision with the TB, is seen to be very sensitive to the soliton power level due the cross-phase modulation among various fission products, especially in the vicinity of the power levels at which the TBS can be generated. The refractive index jump is given by $\Delta n=-0.787\times 10^{-7}$.

\hspace*{\fill} \\
{\bf Supplementary Movie 2:} In Movie 2, we zoom in onto short subranges of the main power range (210W to 290W). Movie 2(a) is a fragment of Movie 1 corresponding to the subrange (210W, 230W) of the peak power of the input pulse. No TBS forms in this subrange and the usual fundamental bulk solitons, resulted from the break-up of the incident soliton upon its collision with the TB, are seen to propagate away from the boundary. Movie 2(b) is a fragment of Movie 1, corresponding to the input peak power subrange (240W, 260W). A TBS can only form at specific power values within this power range, and the soliton fission scenario is highly sensitive to the input power level.

\hspace*{\fill} \\
{\bf Supplementary Movie 3:} This movie shows the evolution of the pulse propagating in a nonlinear optical medium  as the input peak power $P_0$ is varied from 644W to 744W. Notice that only in very rare cases can one- and two-TBS states form. Observe that it requires higher power levels to generate two TBSs within a common self-induced waveguide by the TB than it does to trap a single TBS there. The refractive index jump is given by $\Delta n=-0.804\times 10^{-7}$.

\hspace*{\fill} \\
{\bf Supplementary Movie 4:} To illustrate the TBS emergence in greater detail, we present several fragments of Movie 3 by focusing on short subranges of the main power range (644W, 744W). Movie 4(a) is a fragment corresponding to the input peak power subrange (692.5W, 694.5W). No TBS emerges in this power subrange. Movie 4(b) is a fragment corresponding to the input peak power subrange (694.5W, 696.5W). The soliton fission scenario at the TB in this power subrange involves a one-TBS state formation and is very sensitive to the input power level due to the cross-phase modulation. Movie 4(c) corresponds to the input peak power subrange (715.8W,717.8W); it exhibits two-TBS state formation and extreme sensitivity of the soliton fission scenario to the input power level. Movie 4(d), corresponding to the power subrange (719W,721W), reveals no TBS formation. The video reiterates our key message: the TBS can only emerge under very special circumstances---corresponding to specific values of the peak power of the input pulse---and only extremely rarely can the TBS arrive to the observation plane $z=z_{\ast}$ intact.

\end{document}